# Effect of Surfaces on Amyloid Fibril Formation

Bradley Moores[1], Elizabeth Drolle[2], Simon J. Attwood[1], Janet Simons[2], Zoya Leonenko[1,2]*

1 Department of Physics and Astronomy, University of Waterloo, Waterloo, Ontario, Canada, 2 Department of Biology, University of Waterloo, Waterloo, Ontario, Canada

**Abstract**

Using atomic force microscopy (AFM) we investigated the interaction of amyloid beta (Aβ) (1–42) peptide with chemically modified surfaces in order to better understand the mechanism of amyloid toxicity, which involves interaction of amyloid with cell membrane surfaces. We compared the structure and density of Aβ fibrils on positively and negatively charged as well as hydrophobic chemically-modified surfaces at physiologically relevant conditions. We report that due to the complex distribution of charge and hydrophobicity amyloid oligomers bind to all types of surfaces investigated ($CH_3$, COOH, and $NH_2$) although the charge and hydrophobicity of surfaces affected the structure and size of amyloid deposits as well as surface coverage. Hydrophobic surfaces promote formation of spherical amorphous clusters, while charged surfaces promote protofibril formation. We used the nonlinear Poisson-Boltzmann equation (PBE) approach to analyze the electrostatic interactions of amyloid monomers and oligomers with modified surfaces to complement our AFM data.

**Citation:** Moores B, Drolle E, Attwood SJ, Simons J, Leonenko Z (2011) Effect of Surfaces on Amyloid Fibril Formation. PLoS ONE 6(10): e25954. doi:10.1371/journal.pone.0025954

**Editor:** Igor Sokolov, Clarkson University, United States of America

**Received** June 7, 2011; **Accepted** September 14, 2011; **Published** October 10, 2011

**Copyright:** © 2011 Moores et al. This is an open-access article distributed under the terms of the Creative Commons Attribution License, which permits unrestricted use, distribution, and reproduction in any medium, provided the original author and source are credited.

**Funding:** This work was supported by the Natural Science and Engineering Council of Canada (NSERC), http://www.nserc-crsng.gc.ca, Canada Foundation for Innovation, www.innovation.ca, and Ontario Ministry of Research and Innovation, www.mri.gov.on.ca: Ontario Research Fund (ORF), and Waterloo Institute for Nanotechnology (WIN), www.nano.uwaterloo.ca. JS is an undergraduate student supported by NSERC USRA award, BM is a MS student supported by NSERC Discovery grant (ZL) and WIN Graduate Fellowship (BM), ED is a PhD student supported by NSERC Graduate Scholarship (ED), Simon Attwood is a Postdoctoral Fellow supported by NSERC Discovery grant (ZL). The funders had no role in study design, data collection and analysis, decision to publish or preparation of the manuscript.

**Competing Interests:** The authors have declared that no competing interests exist.

* E-mail: zleonenk@uwaterloo.ca

## Introduction

Amyloid fibrils are implicated in many neurodegenerative diseases for which no cure is currently available, including Alzheimer's (AD), Huntington's and Parkinson's diseases [1–5]. Despite the differences in the native structures and functions of the amyloid forming proteins, they form similar fibrils irrespective of the protein from which they originate [4,5]. The molecular mechanism of amyloid toxicity is not well understood. The proposed mechanism of amyloid fibril formation involves protein cleavage from the membrane, unfolding and formation of amyloid fibrils [6]. Although fibril plaque formation is associated with biological membranes in vivo, the role of membrane surfaces is not well understood. A growing number of recent research contributions suggest the importance of membrane surfaces [4,7–16] and in particular the role of electrostatic interactions between the lipid membrane and amyloid-forming proteins [17–19]. Therefore, the detailed investigation of amyloid fibril formation on the surface of lipid membrane is extremely important and may provide an insight into understanding the mechanism of amyloid fibril formation and toxicity. While biological surfaces are extremely important in protein adsorption and amyloid fibril formation, interpreting the results of heterogeneous and complex systems like plasma membranes is often very difficult. Chemically modified surfaces with well-defined physical properties can be considered as simplified models to study the effect of surfaces on amyloid binding and fibril formation. The study of protein aggregation on surfaces has recently attracted a lot of attention [12,20–24] and grown even more important due to the increased use of inorganic and synthetic surfaces as interfaces in bio- and nano-technology.

It has become evident that surfaces play a crucial role in amyloid fibril formation for many amyloidogenic peptides. The size and shape of amyloid aggregates and fibrils, as well as the kinetics of their formation are affected by the physicochemical nature of the surface. It has been shown that for many amyloid peptides, fibril formation is accelerated significantly by surfaces when compared to fibrillization in solution [24,25]. In addition to catalyzing the rate of fibril formation, the mechanism of fibrillization on surfaces has been shown to be different from that in solution [24].

Although it has been shown that the surfaces play an important role in amyloid fibril formation, electrostatic interaction cannot be easily compared as often experiments presented by different research groups are done at varying experimental conditions, such as pH, temperature, the type of surfaces and the type of peptide used. Protein binding to surfaces at high temperatures cannot be compared to experiments on surfaces conducted at room temperatures [26], nor can different experiments done at different pH [26–31] be compared to elucidate the effect of surfaces. This comparison is also difficult due to the limited type of surfaces used [24] for the same type of amyloid proteins [25,26]. Therefore, more work is required to understand the effect of surface functionality using simple model surfaces before moving to more complex surfaces of lipid bilayer or plasma membrane. Our hypothesis is that the surface charge and hydrophobicity affects the structure, amount and surface coverage of Aβ deposits and may play an important role when interactions of Aβ with cell surfaces are considered. To make a clear comparison these surfaces need to be compared under the same conditions in order to elucidate their effect on Aβ aggregation. Recently Wang et al [32,33] reported a systematic molecular dynamics study on Aβ binding to self-assembled thiol monolayers with four different functional groups.

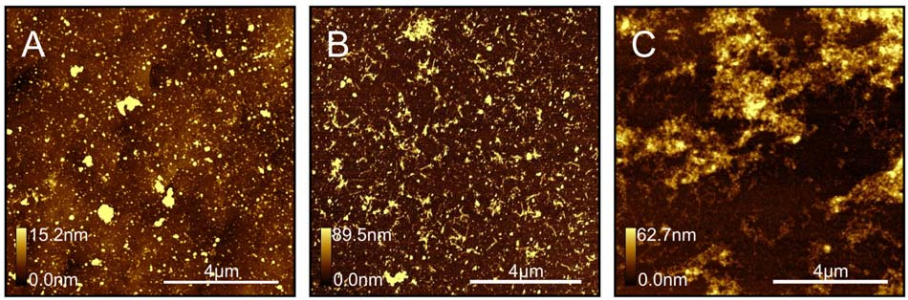

Figure 1. AFM topography images (10×10 μm) of amyloid fibril formation on $NH_2$- modified surfaces: (A) after 10 minutes, (B) after 6 hours, and (C) after 22 hours incubation at 37°C in HEPES buffer, pH 7.8.
doi:10.1371/journal.pone.0025954.g001

In our experiments using high resolution atomic force microscopy we studied the interaction of Aβ (1–42) with three different surfaces: positively charged ($NH_2$), negatively charged (COOH) and hydrophobic ($CH_3$) modified surfaces at pH 7.8, at 37°C, in order to determine the effect of these surfaces on amyloid aggregation and fibril formation. In order to understand electrostatic interactions of amyloid aggregates with functionalized surfaces we employed the nonlinear Poisson-Boltzmann equation (PBE) approach [34]. Using this methodology [34–38], it has been demonstrated recently that electrostatic potentials can be successfully calculated for large micromolecules and bioassemblies, which helps to understand the function of these structures. We used the PBE approach to analyze the interactions of amyloid monomers and oligomers with thiol-modified surfaces and compared these to AFM data. To the best of our knowledge, this is a first report where electrostatic interactions of Aβ peptide and oligomers were compared in similar experimental conditions, combining experimental AFM data and PBE theoretical analysis.

## Results and Discussion

Amyloid plaques are formed when proteins which exist in an alpha-helical form, unfold, convert to beta-sheets and form fibrils. Amyloid fibril formation has been studied extensively in solution where the interaction between peptide molecules is mainly considered. According to the proposed hypothesis, interaction between amyloidogenic peptides in solution may result in the formation of various aggregates, such as small oligomers and long fibrils with twisted morphology [3,39] which are formed by attaching monomer units to the end of growing fibrils. However, the structure of protofibrils formed on the surface is different from the twisted morphology of fibrils formed in solution [25,40,41]. The proposed mechanism includes two distinct stages: nucleation and elongation [25,40–44]. According to this model, aggregation of amyloid peptides on surfaces occurs via formation of small oligomeric units. Bidirectional elongation of so-called protofibrils on the surface occurs with the addition of monomers, oligomeric building blocks, or smaller protofibrils. Investigation of these oligomers is of high importance as they recently have been recognized to interact actively with the cell surfaces inducing more toxic effects [45–48] than mature Aβ fibrils, as it was assumed earlier [49–52]. The size of these oligomeric units depends on the type of the protein and may correspond to monomer, dimer, or small oligomer, or, in some cases, may include up to 20–100 individual peptide molecules [25,40,44].

### Time Dependence.

We investigated fibril formation of Aβ (1–42) peptide on chemically modified surfaces bearing $CH_3$, COOH, and $NH_2$ functional groups. A progressive accumulation of Aβ deposits with time was observed on all surfaces. Figure 1 shows the increase of amyloid fibril formation with time on an $NH_2$-modified surface. At 10 minutes incubation (Figure 1A) mostly small spherical aggregates and a few small protofibrils were observed on the surface. After 6 hours incubation (Figure 1B), both small spherical aggregates and a growing number of protofibrils were observed. After 22 hours incubation (Figure 1C), fibrils grow in size and form larger clusters of longer fibrils and small oligomers. These amyloid clusters vary from 20–70 nm in height, and fibrils are 500 nm to 2 μm long, 4 nm or 6–8 nm high. Underneath the larger clusters, the surface is covered with small spherical aggregates approximately 2 nm high. These small spherical aggregates were visible on the surfaces at all incubation times. A similar increase in size and amount of aggregation was observed on two other functionalized surfaces, as shown in Figure 1.

### Effect of Surfaces

Although all surfaces promoted adsorption of similar smaller building blocks, we observed that the $CH_3$-modified surface promoted formation of amorphous aggregates, while hydrophilic $NH_2$- and COOH-modified surfaces showed clusters of small spheres and short protofibrils (Figure 2). Figure 2 shows the AFM topography images of amyloid deposits formed after 22 hours incubation on $CH_3$-terminated surface (Figure 2A), $NH_2$-terminated surface (Figure 2B), and COOH-terminated surface (Figure 2C).

Each of these surfaces after 22 hours incubation show large clusters of Aβ aggregates. The hydrophobic $CH_3$-terminated surface (Figure 2A) shows amorphous globular clusters of various sizes joined together, but no long separated fibrils were observed on this surface. Both the positively charged $NH_2$-surface (Figure 2B) and negatively charged COOH-surface (Figure 2C) show clusters of fibril-like structures and uniformly sized globular aggregates. We found that smaller spherical aggregates cover the surface uniformly in between larger clusters (Figure 2 B and C), forming a monolayer which covers the surface completely underneath larger clusters. The density of larger clusters is highest on $CH_3$-modified surface (Figure 2A) and lowest on $NH_2$-modified surface (Figure 2B). We performed a statistical analysis of aggregate surface coverage at 22 hours incubation, counting for the second layer of amyloid deposits formed. The $CH_3$-modified surface is covered by amyloid deposits almost completely (94% surface coverage), whereas the COOH- and $NH_2$-modified are surfaces covered by amyloid deposits in a less extent, 81.3% and 23.7% respectively (Figure 2).

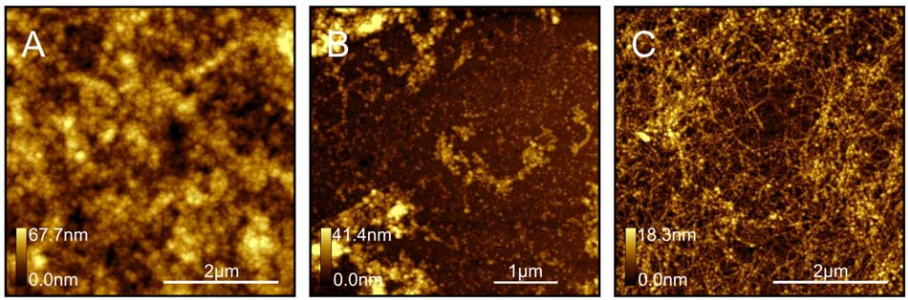

Figure 2. AFM topography images (5×5 μm) of the amyloid fibrils formed on CH$_3$, NH$_2$, and COOH –modified surfaces. Aβ peptide solution was incubated for 22 hours at 37°C on: (A) CH$_3$-, (B) NH$_2$-, and (C) COOH-modified surfaces.
doi:10.1371/journal.pone.0025954.g002

High resolution images, Figure 3, show clearly that the first monolayer on each surface type is composed of smaller aggregates densely packed together. This monolayer is formed after 1 hour incubation. At 1 hour incubation we also can clearly see few small and separated protofibrils, formed on the top of the monolayer, Figure 3 A, B, C, D, and G. These protofibrils are also composed of spherical aggregates, as shown on Figure 3C, similar in size to the monolayer components.

The shape and size of small oligomers were slightly different for different surfaces. Figure 3C, 3F, and 3I show small scan areas of the monolayer for each surface type. Figure 4 shows results of statistical analysis of size distribution. Both surface area and height plots indicate that the negatively charged COOH-modified surface (Figure 4 C and D) and positively charged NH$_2$-modified surface (Figure 4 E and F) have smaller aggregates than the CH$_3$–modified surface (Figure 4 A and B). The Aβ oligomers on the hydrophobic CH$_3$–modified surface are mostly spherical and less uniform in size, which is revealed by broad

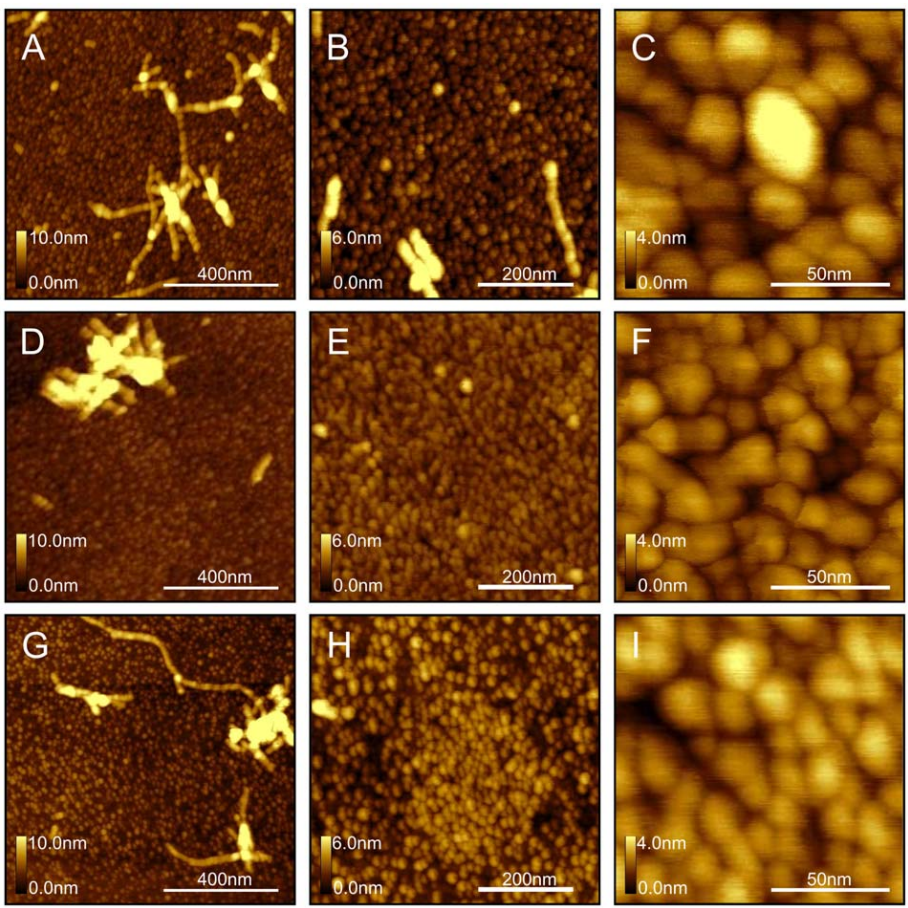

Figure 3. High resolution images of AFM topography of Aβ aggregates formed on modified surfaces: CH$_3$- (A–C), COOH- (D–F), and NH$_2$- (G–I) modified surfaces, after incubation with Aβ (1–42) solution (500 μg/ml) for 1 hour at 37°C.
doi:10.1371/journal.pone.0025954.g003

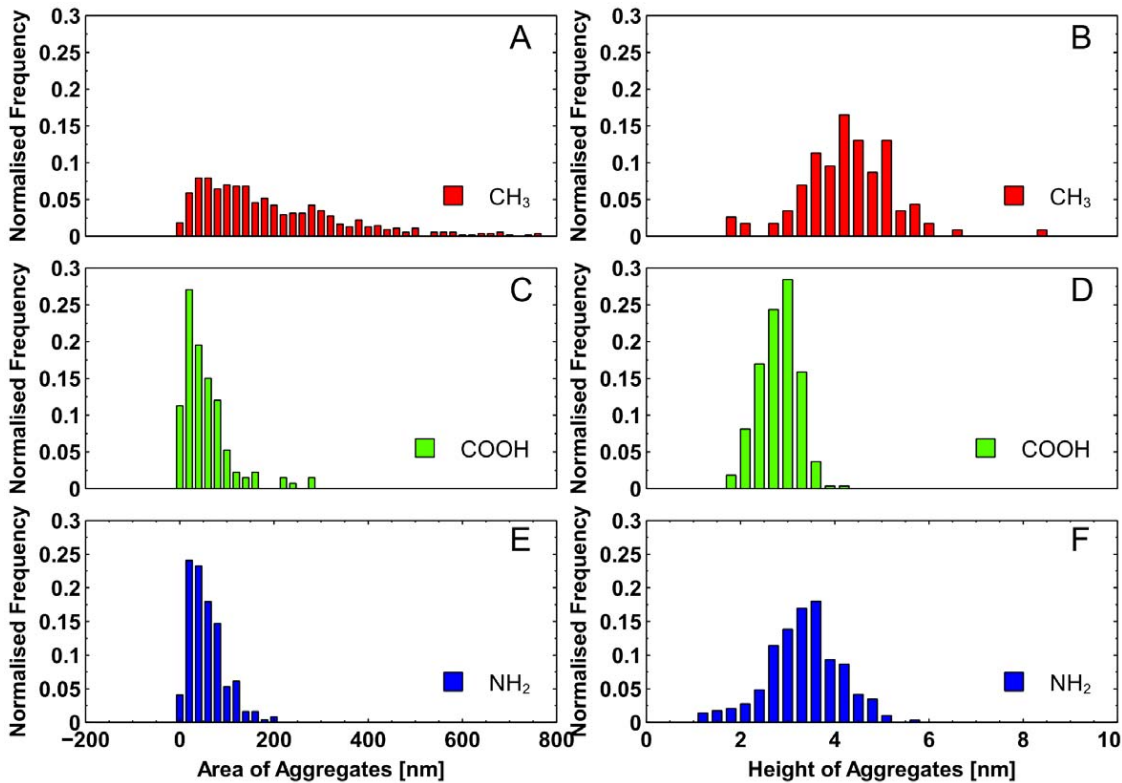

**Figure 4. Statistical analysis of Aß small aggregates shown in figure 3.** Histograms of aggregate unit area size (A, C, E) and height of each aggregate (B, D, F) for each surface type; (Red) $CH_3$-modified surface; (green) COOH-modified surface; (blue) $NH_2$-modified surface.
doi:10.1371/journal.pone.0025954.g004

height distribution, Figure 4, compared to deposits on charged $NH_2$ and COOH surfaces.

The presence of small oligomeric units correlates with previously reported oligomer building blocks [39] for Aβ (1–40). The authors [39–41] observed spherical units of Aβ (1–40) of varying dimensions immediately following the initiation of fibrillization. It has been shown that stable oligomers of a range of molecular weights of both Aβ (1–40) and Aβ (1–42) were isolated from brain and synthetic amyloid material. Size-exclusion chromatography of Aβ deposits has previously revealed dimeric (9 kDa) and trimeric (13.5 kDa) forms [39,53] whereas incubation of monomeric Aβ has led to the separation of 4-, 19-, and 46-kDa fragments [54]. Others also reported the presence of oligomer units [53,55,56].

Although we observed small oligomers present on all types of surfaces used, interestingly, these smaller building blocks on the COOH-terminated surface were not spherical, but rather triangular. This may be a result of the interaction with the negatively charged COOH surface. The aggregates making up the monolayer on the $NH_2$ surfaces were also not completely spherical, but resemble a triangular shape. They appear to be a similar shape to the COOH surface, but more tightly packed together. This indicates that electrostatic interactions with the surfaces may affect the oligomer folding and packing and therefore the shape of the smaller building blocks. Small size triangular shaped oligomers were not observed before, but were proposed by simulations of trimer structures by Paravastu et al. 2008 [57] versus spherical – dimer structures. Wang et al [32] also has shown by molecular dynamics simulations that Aβ is relatively free to move at the $CH_3$ surfaces but stick to COOH- and $NH_2$-surfaces, which may result in more ordered appearance of Aβ deposits on charged surfaces, compared to $CH_3$-surfaces [32].

The electrostatic interactions between the charged surfaces and Aβ directly influence the structure of formed amyloid deposits and may affect the secondary structure of amyloid in these clusters. These electrostatic interactions can be understood when we consider complex charge distribution in Aβ peptide and its dependence on secondary structure. There are six negatively charged residues and three positively charged residues in the peptide, yielding a net charge of -3, with isoelectric point of about 5.5 [58]. Figure 5 demonstrates the organization of charge within various peptide secondary structures. For the α-helix structure (Figure 5A), the charge is fairly evenly distributed to prevent a dipole from forming. In the case of a β-sheet (Figure 5B), a strong positively charged region (blue) forms on either side of peptide, and the negatively charged (red) regions are dispersed through the remainder of the peptide. However, when several β-sheets are stacked together (Figure 5C), strong charged regions form within the aggregate creating a quadrupole moment. Therefore, based on this analysis, we expect that α-helical peptides preferentially form on the hydrophobic $CH_3$ surface, and β-sheet clusters of various sizes on the negatively charged COOH and positively charged $NH_2$ surfaces.

Our findings at neutral pH correlate with work by McMasters et al [27], where the authors investigated amyloid fibril formation of Aβ peptide on chemically modified mica bearing positively or negatively charged, or hydrophobic functional groups at pH 11.5. Using reflection-absorption infrared spectroscopy, the authors found that surfaces covered with sulfonic acid, carboxylic acid, alcohol, and trifluoro-terminated thiol monolayers all cause adsorption/deposition of Aβ (10–35) peptide. Deposits were composed of peptides in beta-sheet, beta-turn, random coil and α-helical conformation. The $CF_3$ monolayer study revealed that equilibrium is slightly shifted towards an α-helix form. This correlates with our results showing amorphous aggregates, in α-helix form and no fibrils on the $CH_3$-modified surfaces. Our observation is consistent also with findings by Giacomelli et al [28], who used spectroscopy methods and showed that adsorption of Aβ (1–40) (at pH 7 and 10, at 25°C) on both hydrophobic Teflon and hydrophilic silica solid surfaces causes conformational changes of the adsorbed peptide, inhibiting polymerization, which occurs in solution during incubation. The conformation of the peptide strongly depends on the hydrophobicity of the surface: hydrophobic interactions promote intramolecular α-helix formation, whereas electrostatic interactions promote intermolecular β-sheet formation. In addition to different structures, we showed that surface-mediated aggregation occurs faster for $CH_3$-modified and COOH-modified surfaces as compared to $NH_2$-modified surfaces. This is indicated by larger amyloid surface coverage for $CH_3$-modified surface and COOH-modified surface shown on Figure 2.

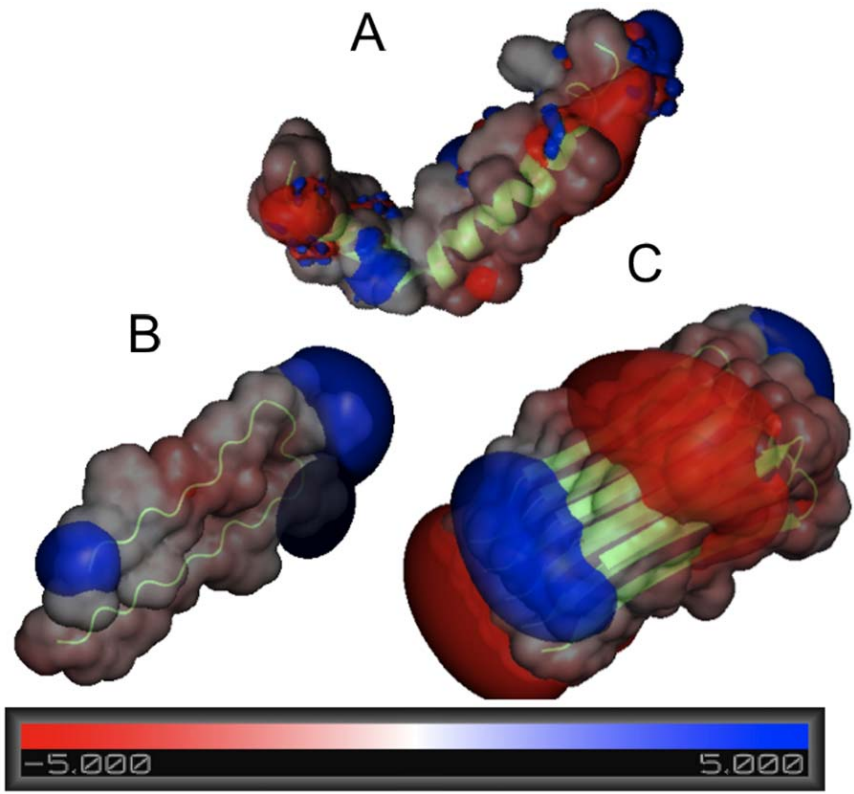

**Figure 5. Electrostatic potentials of amyloid monomers and oligomers.** The 5kT and -5kT isoelectric potential surfaces are superimposed on the molecular surface. Positive charge is shown in blue, and negative in red. The molecular surface is produced by convolving a 1.4Å sphere (which represents a water molecule) around each peptide. The (A) alpha helix monomer does not have any strongly charged regions, whereas the (B) beta sheet monomer has a strong positively charged end. In comparison, the (C) stack of 5 beta sheets has 4 strongly charged regions, which would greatly contribute to electrostatic interactions with charged surfaces. Images were produced using PyMOL v1.2.
doi:10.1371/journal.pone.0025954.g005

In addition our analysis of electrostatic potential distribution using PBE shows that surface charge distribution is different in Aβ monomer, dimer or larger oligomers. Oligomers in β-sheet conformations show larger collective polarity, which induces stronger electrostatic interactions with surfaces, as well as preferential ordering on the oligomers on the surfaces. This may be the driving force for more ordered and fibril-like structures observed on charged surfaces, compared to $CH_3$-modified surfaces. Electrostatic forces induced by surfaces may also drive the re-distribution of electrostatic potential in monomers near the surfaces and therefore may change the secondary structure of the peptide, thus inducing electrostatically driven amyloid fibril formation.

### Incubation of Amyloid Peptide in Solution

We observed that protofibrils formed on surfaces are composed of small spherical units, branched and can grow in any direction by adding the spherical building blocks. Unlike protofibrils formed on surfaces (Figure 3) which are composed of small spheres, fibrils formed in solution (Figure 6) are long, continuous and twisted together into helices, and do not reveal any bead-like structure. This correlates with previously reported data by Blackley et al [40]. The smaller spherical building blocks were not commonly observed in our experiments for fibrils formed in solution (Figure 6), and we rarely observed twisting of protofibrils formed on the surfaces. This suggests that the unfolding of the oligomer units in order to form twisted fibrils is hindered by the surfaces. In solution the peptides have more degrees of freedom (such as rotational, translational, and protein folding), whereas on the surface the degrees of freedom are significantly limited. Additionally, once a peptide is bound to the surface, the peptide has likely found an energy minimum, and therefore requires an energy input to overcome the potential well. Gibbs free energy decreases as a result of the protein absorbing to the surface, and therefore energy must be given to the protein to overcome this decrease [59]. This is consistent with the molecular dynamics simulations data [60,61], which indicate that for larger oligomer sizes or long chain lengths, it is very unlikely for the chain to fold into ordered β-sheet structures. It is much more common for the chains to fold into amorphous aggregates which are in dynamic equilibrium.

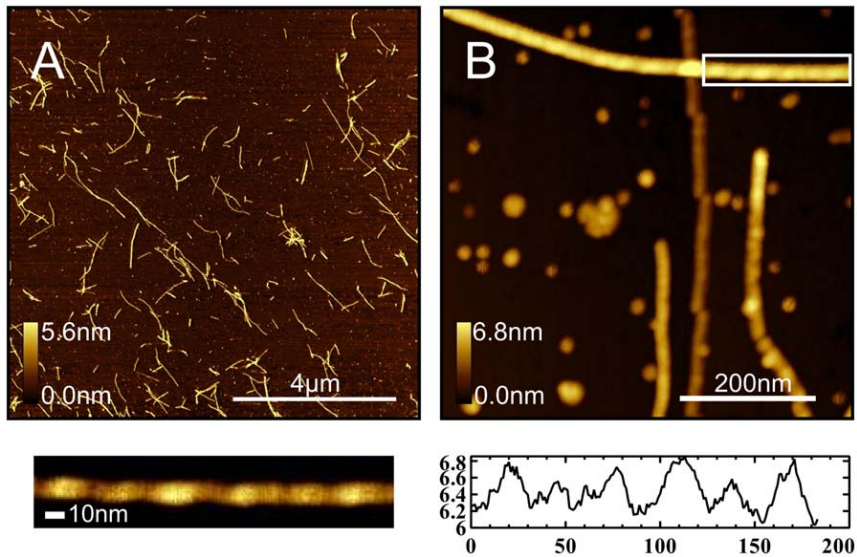

**Figure 6. AFM topography images of Aβ (1–42) incubated in solution (500 μg/mL concentration) at 37°C for 22 hours.** After incubation, 10 μL of solution was deposited onto cleaved mica for 5 minutes, followed by rinsing with nanopure water and drying with a gentle nitrogen stream. A 10×10 μm scan area (A) of the surface shows many fibrils ranging from 0.1 – 4 μm. (B) High resolution image (500×500nm) demonstrates that fibrils reveal twisted morphologies, characteristic of formed in solution; (below, left) expanded view of region enclosed by white box in (B) with optimized z-scale; (below, right) height profile along fibril axis clearly demonstrating maxima and minima which is characteristic of the twisted morphology.
doi:10.1371/journal.pone.0025954.g006

## Conclusions

We investigated the interaction of Aβ (1–42) peptide with three different chemically modified surfaces and compared the effect of these surfaces at the same physiologically relevant conditions. We found that due to the complex dipole distribution amyloid oligomers bind to all surface types investigated (-$CH_3$, -COOH, and -$NH_2$) although the size and shape of these amyloid deposits depend on surface properties. Hydrophilic surfaces show protofibrils coexisting with spherical oligomer aggregates, while hydrophobic $CH_3$-modified surfaces cause formation of amorphous spherical aggregates. The surface charge and hydrophobicity define both the structure of the fibril aggregates formed on the surfaces and kinetics of their accumulation. In addition our analysis of electrostatic potential distribution using PBE shows that surface charge distribution changes depending on the secondary structure of the peptide and may play an important role in electrostatically driven amyloid fibril formation on surfaces.

## Materials and Methods

### Chemicals and Sample Preparation

Decanethiol, 3-mercaptoethanol, APTES (3-Aminopropyl-triethoxysilane) and HEPES buffer were purchased from Aldrich Chemical Co. HPLC grade ethanol was purchased from Sigma. All chemicals were used as received. Water used for sample preparation was purified (distilled de-ionized, millipore water).

### Substrate Preparation

Atomically flat gold on mica surfaces were purchased from Agilent Technologies, Inc. (Santa Clara, CA). These gold surfaces were affixed to clean glass cover slips using Epo-Tek 377 glue from Epo-Tek, Inc. (Billerica, MA), which was cured at 150°C for 1 hour. The gold glued to the glass was peeled from the mica, revealing the atomically flat gold surface. The gold surfaces were further modified by incubating in an appropriate 5 mM thiol solution in ethanol for 48 hours. Prepared mica-gold substrates were modified with decanethiol and 3-mercaptoethanol, and pure glass substrates were modified with APTES.

### Amyloid Peptide Preparation and Incubation

Aβ (1–42) was purchased freeze-dried in 0.5 mg vials from rPeptide (Atlanta, GA). These amyloid samples were pretreated according to the Fezoui et al. (2000) procedure [62] to ensure monomeric solution. After this each 0.5 mg aliquot was dissolved in 1 mL of pH 7.8 50 mM HEPES buffer. Small amount of the protein solution (50 μL aliquots) were immediately placed on modified surface and were incubated at 37°C in liquid cell. After defined period of time (from 10 min to 22 hours) samples were rinsed with millipore DI water, dried with gentle stream of nitrogen, and kept in a desiccator prior to imaging.

### AFM Imaging and Analysis

Imaging was done in intermittent contact mode on a JPK Nanowizard II atomic force microscope (AFM) recorded with AC mode, and on an Agilent AFM/SPM-5500 AFM using MAC mode imaging. Images were obtained with Nanoworld NCH tips, with a resonant frequency of 338 kHz and 42 N/m spring constant in air or Agilent MAC mode cantilevers, with a resonant frequency of 75 kHz and a spring constant of 2.8 N/m in air.

Statistical analysis was done with the program called CellProfiler. The shape of each amyloid aggregate was measured using image recognition. The purpose of the program is to count the number of spherical aggregates (or cells) in an image. Using this program we determined the size and shape of the aggregates. The total number of measurements used for analysis is: 133 for $CH_3$, 543 for COOH, and 245 for $NH_2$ surfaces.

### Electrostatic Modeling

Electrostatic potentials of Aβ monomers in β-sheet and α-helix conformations, and amyloid oligomers were obtained by solving the Adaptive Poisson-Boltzmann Solver (APBS v1.2.1b) [35,63] implementation for PyMol v1.2. Briefly, the APBS method uses the finite element method to numerically solve the nonlinear Poisson-Boltzmann equation, representing the electrostatic interactions between molecules in aqueous environments [34]. We constructed surface potential profiles for Aβ monomers in β-sheet and α-helix conformations and amyloid pentamers based on crystalline structure obtained from protein data bank [64,65].

The electrostatic interaction energies are calculated for each voxel within a defined volume. In our calculations the solvent dielectric constant was 80, and the protein dielectric was approximated as 2.0. The isoelectric potential of 5kT and -5kT was mapped onto the A-β monomers and beta sheet oligomer. The monomers and oligomers are displayed by convolving a 1.4Á sphere representing the approximate solvent size.

### Acknowledgments

The authors acknowledge technical support from Dr. Song Xu (Agilent Technologies) and Dr. Christian Löbbe, Dr. Sid Ragona (JPK Instruments).

### Author Contributions

Conceived and designed the experiments: ZL BM. Performed the experiments: BM JS. Analyzed the data: BM ED SA. Contributed reagents/materials/analysis tools: ZL. Wrote the paper: BM JS ED SA ZL.

### References

1. Lashuel HA, Lai Z, Kelly JW (1998) Characterization of the Transthyretin Acid Denaturation Pathway by Analytical Ultracentrifugation: Implications for wild type, V30M and L55P Amyloid Fibril Formation. Biochem 37: 17851–17864.
2. Lansbury P, Rochet J (2000) Amyloid fibrillogenesis: themes and variations. Curr Opin Struct Biol 10: 60–68.
3. Chiti F, Dobson CM (2006) Protein misfolding, functional amyloid, and human disease. Annu Rev Biochem 75: 333–366.
4. Dobson CM (2001) The Structural Basis of Protein Folding and its Links with Human Disease. Phil Trans R Soc Lond B 356: 133–145.
5. Johansson J (2003) Molecular determinants for amyloid fibril formation: lessons from lung surfactant protein C. Swiss Med Wkly 133: 275–282.
6. Carrell R, Gooptu B (1998) Conformational Changes and Disease - Serpins, Prions and Alzheimer's. Curr Opin Struct Biol 8: 799–809.
7. Lin HAI, Bhatia R, Lal R (2001) Amyloid Beta Protein Forms Ion Channels: Implications for Alzheimer's Disease Pathophysiology. FASEB J 13: 2433–2444.
8. Quist A, Doudevski I, Lin H, Azimova R, Ng D, et al. (2005) Amyloid Ion Channels: A Common Structural Link for Protein-Misfolding Disease. Proc Natl Acad Sci USA 102: 10427–10432.
9. Kayed R, Sokolov Y, Edmonds B, McIntire TM, Milton SC, et al. (2004) Permeabilization of Lipid Bilayers is a Common Conformation-Depedent Activity of Soluble Amyloid Oligomers in Protein Misfoldong Diseases. J Biol.Chem 279: 46363–46366.
10. Ambroggio EE, Kim DH, Separovic F, Barrow CJ, Barnham KJ, et al. (2005) Surface Behavior and Lipid Interaction of Alzheimer Beta-Amyloid Peptide 1-42: A Membrane-Disrupting Peptide. Biophys J 88: 2706–2713.
11. Cordy JM, Hooper NM, Turner AJ (2006) The Involvement of Lipid Rafts in Alzheimer's Disease. Mol Membr Biol 23: 111–122.
12. Sharp JS, Jones RAL, Forrest JA (2002) Surface Denaturation and Amyloid Fibril Formation of Insulin at Model Lipid-Water Interfaces. Biochem 41: 15810–15819.
13. Ege C, Majewski J, Wu G, Kjaer K, Lee KY (2005) Templating Effect of Lipid Membranes on Alzheimer's Amyloid. Chemphyschem 6: 226–229.
14. Yip CM, Elton EA, Darable AA, Morrison MR, McLaurin J (2001) Cholesterol, A Modulator of Membrane-Associated Aβ-Fibrillogenesis and Neurotoxicity. J Mol Biol 311: 723–734.
15. Choucair A, Chakrapani M, Chakravarthy B, Katsaras J, Johnston L (2007) Preferential Accumulation of Abeta(1-42) on Gel Phase Domains of Lipid Bilayers: An AFM and Fluorescence Study. Biochim Biophys Acta Biomembr 1768: 146–154.
16. Terzi E, Hölzemann G, Seelig J (1995) Self-Association of Beta-Amyloid Peptide (1-40) in Solution and Binding to Lipid Membranes. J Mol Biol 252: 633–642.
17. Choo-Smith LP, Rodriguez W, Glabe C, Surewicz W (1997) Acceleration of Amyloid Fibril Formation By Specific Binding of β-Amyloid Peptide (1–40) to Ganglioside-Containing Membrane Vesicles. J Biol Chem 272: 22987–22990.
18. Kakio A, Nishimoto S, Yanagisawa K, Kozutsumi Y, Matsuzaki K (2002) Interactions of Amyloid Beta-Protein with Various Gangliosides in Raft-Like Membranes: Importance of GM1 Ganglioside-Bound Form as an Endogenous Seed for Alzheimer Amyloid. Biochem 41: 7385–7390.
19. Yanagisawa K, Odaka ASuzuki N, Ihara Y (1995) GM1 Ganglioside-Bound Amyloid β-Protein (Aβ): A Possible Form of Preamyloid in Alzheimer's Disease. Nat Med 1: 1062–1066.
20. Nayak A, Dutta AK, Belfort G (2008) Surface-Enhanced Nucleation of Insulin Amyloid Fibrillation. Biochem Biophys Res Comm 369: 303–307.
21. Sethuraman A, Belfort G (2005) Protein Structural Perturbation and Aggregation on Homogeneous Surfaces. Biophys J 88: 1322–1333.
22. Yang H, Fung S, Pritzker M, Chen P (2007) Surface-Assisted Assembly of an Ionic-Complementary Peptide: Controllable Growth of Nanofibers. J Am Chem Soc 129: 12200–12210.
23. Sigal GB, Mrksich M, Whitesides GM (1998) Effects of Surface Wettability on the Adsorption of Proteins and Detergents. J Am Chem Soc 120: 3464–3473.
24. Powers ET, Kelly JW (2001) Medium-Dependent Self-Assembly of an Amphiphilic Peptide: Direct Observation of Peptide Phase Domains at the Air-Water Interface. J Am Chem Soc 123: 775–776.
25. Zhu M, Souillac PO, Ionescu-Zanetti C, Carter SA, Fink AL (2002) Surface-Catalyzed Amyloid Fibril Formation. J Biol Chem 277: 50914–50922.
26. Hammarström P, Ali MM, Mishra R, Svensson SA (2008) Catalytic Surface for Amyloid Fibril Formation. J Phy. Conf Ser 100: 052039.
27. McMasters MJ, Hammer RP, McCarley RL (2005) Surface-Induced Aggregation of Beta Amyloid Peptide by ω-Substituted Alkanethoil Monolayers Supported on Gold. Langmuir 21: 4464–4470.
28. Giacomelli CE, Norde W (2005) Conformational Changes of the Amyloid ß-Peptide (1-40) Adsorbed on Solid Surfaces. Macromol Biosci 5: 401–407.
29. Uversky VN, Li J, Fink AL (2001) Evidence for a Partially-Folded Intermediate in Alpha-Synuclein Fibril Formation. J Biol Chem 276: 10737–10744.
30. Qin Z, Hu D, Zhu M, Fink AL (2007) Structural Characterization of the Partially Folded Intermediates of an Immunoglobulin Light Chain Leading to Amyloid Fibrillation and Amorphous Aggregation. Biochemistry 46: 3521–3531.
31. Yokoyama K, Welchons DR (2007) The Conjugation of Amyloid Beta Protein on the Gold Colloidal Nanoparticles' Surfaces. Nanotechnology 18: 105101–105107.
32. Wang Q, Zhao J, Yu Z, Zhao C, Li L, et al. (2010) Alzheimer A-b 1-42 Monomer Adsorbed on the Self-Assembled Monolayers. Langmuir 26: 12722–12732.
33. Wang Q, Zhao C, Zhao J, Wang J, Yang JC, et al. (2010) Comparative Molecular Dynamics Study of Aβ Adsorption on the Self-Assembled Monolayers, Langmuir 26: 3308–3316.
34. Baker NA, Sept D, Joseph S, Holst MJ, McCammon JA (2001) Electrostatics of Nanosystems: Application to Microtubules and the Ribosome. Proc Natl Acad Sci USA 98: 10037–10041.
35. Sharp KA (1994) Electrostatic Interactions in Macromolecules. Curr Opin Struct Biol 4: 234–239.
36. Guo T, Gong LC, Sui SF (2010) An Electrostatically Preferred Lateral Orientation of SNARE Complex Suggests Novel Mechanisms for Driving Membrane Fusion. PLoS ONE 5: e8900.
37. Dolinsky TJ, Nielsen JE, McCammon JA, Baker NA (2004) PDB2PQR: An Automated Pipeline for the Setup of Poisson-Boltzmann Electrostatics Calculations. Nucleic Acids Res 32: 665–667.
38. Cohen FS, Melikyan GB (2004) The Energetics of Membrane Fusion from Binding, Through Hemifusion, Pore Formation, and Pore Enlargement. J Membr Biol 199: 1–14.
39. Blackley HKL, Patel N, Davies MC, Roberts CJ, Tendler SJB, et al. (1999) Morphological Development of β(1-40) Amyloid Fibrils. Exp Neurol 158: 437–443.
40. Blackley HKL, Sanders GHW, Davies MC, Roberts CJ, Tendler SJB, et al. (2000) In-situ Atomic Force Microscopy Study of β –Amyloid Fibrillization. J Mol Biol 298: 833–840.
41. Rocha S, Krastev R, Thünemann AF, Pereira MC, Möhwald H, et al. (2005) Adsorption of Amyloid β-Peptide at Polymer Surfaces: A Neutron Reflectivity Study. ChemPhysChem 6: 2527–2534.
42. Kowalewski T, Holtzman DM (1999) In situ Atomic Force Microscopy Study of Alzheimer's β-Amyloid Peptide on Different Substrates: New Insights into Mechanism of β-Sheet Formation. Proc Natl Acad Sci USA 96: 3688–3693.
43. Jarrett JT, Berger EP, Lansbury PT (1993) The Carboxy Terminus of the Beta Amyloid Protein is Critical for the Seeding of Amyloid Formation: Implications of the Pathogensis of Alzheimer's Disease. Biochemistry 32: 4693–4697.

44. Lomakin A, Chung DS, Benedek GB, Kirschner DA, Teplow DB (1996) On the Nucleation and Growth of Amyloid β-Protein Fibrils: Detection of Nuclei and Quantitation of Rate Constants. Proc Natl Acad Sci USA 93: 1125–1129.
45. Goldberg MS, Lansbury PT (2000) Is There a Cause-and-Effect Relationship Between Alpha-Synuclein Fibrillizaton and Parkinson's Disease? Nat Cell Biol 2: E115–E119.
46. Stefani M (2007) Generic Cell Dysfunction in Neurodegenerative Disorders: Role of Surfaces in Early Protein Misfolding, Aggregation, and Aggregate Cytotoxicity, Neuroscientist 13: 519–531.
47. Buxbaum J, Gallo G (1999) Nonamyloidotic Monoclonal Immunoglobulin Deposition Disease. Light-Chain, Heavy-Chain, and Light- and Heavy-Chain Deposition Diseases. Hematol Oncol Clin North Am 13: 1235–1248.
48. Lambert MP, Barlow AK, Chromy BA, Edwards C, Freed R, et al. (1998) Diffusible, Nonfibrillar Ligands Derived from Abeta1-42 are Potent Central Nervous System Neurotoxins. Proc Natl Acad Sci USA 95: 6448–6453.
49. Howlett DR, Jennings KH, Lee DCClark MSG, Brown F, et al. (1995) Aggregation State and Neurotoxic Properties of Alzheimer Beta-Amyloid Peptide. Neurodegener 4: 23–32.
50. Pike CJ, Walencewicz AJ, Glabe CG, Cotman CW (1991) In vitro Aging of β-Amyloid Protein Causes Peptide Aggregation and Neurotoxicity. Brain Res 563: 311–314.
51. Pike CJ, Burdick D, Walencewicz AJ, Glabe CG, Cotman CW (1993) Neurodegeneration Induced by β - Amyloid Peptides *in vitro*: The Role of Peptide Assembly State. J Neurosci 13: 1676–1687.
52. Seilheimer B, Bohrmann B, Bondolfi L, Muller F, Stuber D, et al. (1997) The Toxicity of the Alzheimer's Beta-Amyloid Peptide Correlates with a Distinct Fiber Morphology. J Struct Biol 119: 59–71.
53. Roher AE, Chaney MO, Kuo YM, Webster SD, Stine WB, et al. (1996) Morphology and Toxicity of Aβ-(1-42) Dimer Derived from Neuritic and Vascular Amyloid Deposits of Alzheimer's Disease. J Biol Chem 271: 20631–20635.
54. Dyrks T, Dyrks E, Masters C, Beyreuther K (1993) Amyloidogenicity of Rodent and Human Beta A4 Sequences. FEBS Lett 324: 231–236.
55. Kuo Y, Emmerling ER, Vigo-Pelfrey C, Kasunic TC, Kirkpatrick JB, et al. (1996) Water-Soluble Aβ(N-40, N-42) Oligomers in Normal and Alzheimer Disease Brains. J Biol Chem 271: 4077–4081.
56. Levine H (1995) Soluble Multimeric Alzheimer Beta (1-40) Pre-Amyloid Complexes in Dilute Solution. Neurobiol Aging 16: 755–764.
57. Paravastu AK, Leapman RD, Yau WM, Tycko R (2008) Molecular Structural Bias for Polymorphism in Alzheimer's β-Amyloid Fibrils. Proc Natl Acad Sci USA 105: 18349–18354.
58. Rauk A (2009) The Chemsitry of Alzheimer's Disease. Chem Soc Rev 38: 2698–2715.
59. Latour RA, Rini CJ (2002) Theoretical Analysis of Adsorption Thermodynamics for Hydrophobic Peptide Residues on SAM Surfaces of Varying Functionality. J Biomed Mater Res 60: 564–577.
60. Boucher G, Mousseau N, Derreumaux P (2006) Aggregating the Amyloid $A\beta_{11-25}$ Peptide into a Four-Stranded β-Sheet Structure. Proteins: Struct, Funct Bioinf 65: 877–888.
61. Melquiond A, Gelly JC, Mousseau N, Derreumaux P (2007) Probing Amyloid Fibril Formation of the NFGAIL Peptide by Computer Simulations. J. Chem Phys 126: 25101–25106.
62. Fezoui Y, Hartley D, Harper J, Khurana R, Walsh D, et al. (2000) An Improved Method of Preparing the Amyloid P-Protein for Fibrillogenesis and Neurotoxicity Experiments. Amyloid: Int J Exp Clin Invest 7: 166–178.
63. Sharp KA, Honig B (1990) Calulating Total Electrotatic Energies with Nonlinear Poisson-Boltzmann Equation. J Phys Chem 94: 7684–7692.
64. Crescenzi O, Tomaselli S, Guerrini R, Salvadori S, D'Ursi AM, et al. (2002) Solution Structure of the Alzheimer's Disease Amyloid Beta-Peptide (1-42). Eur J Biochem 269: 5642–5648.
65. Luhrs T, Ritter C, Adrian M, Riek-Loher D, Bohrmann, B, et al. (2005) 3D Structure of Alzheimer's Amyloid-β(1-42) Fibrils. Proc Natl Acad Sci USA 102: 17342–17347.